\documentclass[a4paper,12pt]{article}
\usepackage{jcappub}
\pdfoutput=1
\usepackage{graphicx,amsfonts,color,comment,amsmath,float,hyperref}
\usepackage[dvipsnames]{xcolor}
\usepackage{bigints}
\usepackage[normalem]{ulem}
\usepackage{amssymb}
\usepackage{mathrsfs,amssymb}  
\usepackage{cancel}

\definecolor{newgreen}{rgb}{0., 0.56, 0.}

\hypersetup{
    colorlinks=true,
    linkcolor=newgreen,
    filecolor=magenta,      
    urlcolor=purple,
    citecolor=BrickRed,
}

\begin{document}
\title{Time-delayed neutrino emission from supernovae as a probe of dark matter-neutrino interactions}

\author[1,2,3]{Jose Alonso Carpio}
\author[1,2,3]{Ali Kheirandish}
\author[1,2,3,4,5]{Kohta Murase}
\affiliation[1]{Department of Physics, The Pennsylvania State University, University Park, Pennsylvania 16802, USA}
\affiliation[2]{Department of Astronomy and Astrophysics, The Pennsylvania State University, University Park, Pennsylvania 16802, USA}
\affiliation[3]{Center for Multimessenger Astrophysics, Institute for Gravitation and the Cosmos, The Pennsylvania State University, University Park, Pennsylvania 16802, USA}
\affiliation[4]{School of Natural Sciences, Institute for Advanced Study, Princeton, New Jersey 08540, USA}
\affiliation[5]{Center for Gravitational Physics, Yukawa Institute for Theoretical Physics, Kyoto, Kyoto 606-8502, Japan}

\abstract{
Thermal MeV neutrino emission from core-collapse supernovae offers a unique opportunity to probe physics beyond the Standard Model in the neutrino sector.
The next generation of neutrino experiments, such as DUNE and Hyper-Kamiokande, can detect $\mathcal{O}(10^3)$ and $\mathcal{O}(10^4)$ neutrinos in the event of a Galactic supernova, respectively. 
As supernova neutrinos propagate to Earth, they may interact with the local dark matter via hidden mediators and may be delayed with respect to the initial neutrino signal. 
We show that for sub-MeV dark matter, the presence of dark matter-neutrino interactions may lead to neutrino echoes with significant time delays. The absence or presence of this feature in the light curve of MeV neutrinos from a supernova allows us to probe parameter space that has not been explored by dark matter direct detection experiments.}

\notoc
\maketitle

\section{Introduction}
Overwhelming evidence from astronomical and cosmological probes such as galaxy rotation curve measurements \cite{Rubin:1985ze,Begeman:1991iy} and gravitational lensing \cite{1990ApJ...350...23B,Clowe:2006eq,Jee:2007nx,Bradac:2008eu} have shown that there is a significant amount of non-luminous matter, i.e., dark matter (DM), in the Universe. 
However, the particle nature of DM remains unknown \cite{Bertone:2004pz,Feng:2010gw,Bertone:2016nfn,Bernal:2017kxu}. 
Weakly interacting massive particles (WIMPs) have been the leading scenario as the most studied DM candidate. 
Direct and indirect searches for DM have extensively probed the parameter space for WIMPs. More stringent limits on DM have been found from direct or indirect searches, targeting the interaction of WIMPs with Standard Model (SM) particles and signatures from decay or annihilation of DM to SM (e.g.,~\cite{Beacom:2006tt,Yuksel:2007ac,Murase:2012xs,Murase:2012rd,Gaskins:2016cha,Bernal:2017kxu,Roszkowski:2017nbc,Arcadi:2017kky,Arguelles:2019ouk, Kheirandish:2021II}). 
In the meantime, neutrinos have emerged as the key channel in DM searches, especially if DM interacts with the SM particles via the neutrino portal \cite{Lindner:2010rr,GonzalezMacias:2015rxl,Blennow:2019fhy,Hall:2019rld,Hall:2021zsk,Biswas:2021kio,Borah:2021pet}.

Direct detection searches for DM \cite{Aprile:2016swn,Akerib:2016vxi,Wang:2020coa} have not found any evidence for WIMPs. However, they have limited sensitivities to DM masses below 10 GeV. A lower DM mass threshold is achieved for DM-nucleus scattering via the Migdal effect or DM-electron scattering with bound electrons \cite{Essig:2022dfa}. These techniques can probe DM masses down to 1 MeV.  In the meantime, DM phase-space distribution in dwarf spheroidal galaxies suggests that the fermionic DM mass has a lower bound of $\sim 1$ keV \cite{Boyarsky:2008ju}, while the mass range of 1 keV -- 1 MeV is relatively unexplored. Dark matter below an MeV may arise from self-interacting DM freeze-out after neutrino decoupling \cite{Berlin:2017ftj,Berlin:2018ztp} or through DM freeze-in for sufficiently small couplings \cite{Chang:2019xva,Dvorkin:2020xga}. DM self-interactions also reduce the lower bound on DM mass from Lyman-$\alpha$ constraints \cite{Egana-Ugrinovic:2021gnu,Garani:2022yzj}. 

On the other hand, neutrino physics has seen significant progress with standard neutrino oscillation measurements \cite{Acero:2019ksn,Esteban:2020cvm,Abe:2021gky}. Nevertheless, this field has its own unsolved problems, such as the origin of neutrino mass \cite{Farzan:2012sa,Gouvea:2016,Escudero:2016ksa,Escudero:2016tzx} and detector anomalies \cite{LSND:2001aii,Aguilar-Arevalo:2020nvw}.
Additional sterile neutrino states allow an explanation for neutrino masses via the seesaw mechanism, and their existence was also motivated by the LSND and MiniBooNE anomalies. Moreover, sterile neutrinos with masses in the keV range may also be good DM candidates via the seesaw mechanism
\cite{Abada:2014vea,Abada:2014zra,Boulebnane:2017fxw}.

Both active neutrinos states and DM may interact with a new mediator, albeit with different couplings. In the neutrino sector, neutrino self-interactions via these mediators will result in new features such as delaying the free-streaming behavior of neutrinos \cite{Cyr-Racine:2013jua,Archidiacono:2013dua,Lancaster:2017ksf,Oldengott:2017fhy,Kreisch:2019yzn,Blinov:2019gcj,Brinckmann:2020bcn}, which may contribute to the effective number of relativistic species $N_{\rm eff}$ \cite{Escudero:2019gzq,Araki:2021xdk,Huang:2021dba}. 
They could be realized in particle models predicting additional contributions to the muon anomalous magnetic moment, ($g-2)_\mu$, via a gauged $L_\mu$--$L_\tau$ model \cite{He:1990pn,He:1991qd,Ma:2001md,Heeck:2011wj, Araki:2017rex, Carpio:2021jhu}.

Neutrino self-interactions have been constrained through the use of high-energy cosmic neutrinos \cite{Ioka:2014kca,Shoemaker:2015qul,Kelly:2018tyg,Farzan:2018pnk,Murase:2019xqi,Bustamante:2020mep}, cosmological studies \cite{Barenboim:2019tux,Blinov:2019gcj,Forastieri:2019cuf,RoyChoudhury:2020dmd}, accelerator experiments \cite{Arguelles:2018mtc,Bally:2020yid} and laboratory measurements \cite{Laha:2013xua}. For DM, self-interactions \cite{Kouvaris:2014uoa,Bernal:2015ova,Kainulainen:2015sva,Kamada:2016euw,Cirelli:2016rnw,Tulin:2017ara,Kahlhoefer:2017umn,Ren:2018jpt,Hambye:2019tjt,Kaplinghat:2019dhn} were introduced to alleviate problems with the standard cosmological model, such as the ``too big too fail" problem \cite{2011MNRAS.415L..40B}, the ``missing satellite" problem \cite{Klypin:1999uc,Moore:1999nt} and the ``diversity" problem \cite{Oman:2015xda}. 
Neutrino-DM interaction has been extensively considered in the cosmology context \cite{Boehm:2000gq,Boehm:2001hm,Boehm:2004th,Bertschinger:2006nq,Mangano:2006mp,Serra:2009uu,Wilkinson:2014ksa,Aarssen:2012fx,Farzan:2014gza,Boehm:2014vja,Cherry:2014xra,Bertoni:2014mva,Schewtschenko:2014fca}. These interactions may also be used to boost dark matter (see e.g., \cite{Das:2021lcr}). Presence of such interaction would alter the expansion rate of the Universe, which could affect the observables of the big bang nucleosynthesis (BBN) and cosmic microwave background (CMB). In addition, an ongoing neutrino-DM scattering would damp the power spectrum of primordial fluctuations (see e.g., \cite{Boehm:2014vja}). Observation of high-energy cosmic neutrinos \cite{IceCube:2013cdw, Aartsen:2013jdh, Aartsen:2015dlt} has provided further power to probe for new physics.
Nonstandard neutrino interactions have been studied in this context \cite{Ioka:2014kca,Shoemaker:2015qul}, which bestowed competitive limits with cosmological studies. These searches utilize features induced by DM-neutrino interaction in energy spectrum \cite{Choi:2019ixb}, arrival direction \cite{Arguelles:2017atb}, and arrival time \cite{Murase:2019xqi} of high-energy cosmic neutrinos. The latter has become possible with recent progress in the identification of coincident high-energy neutrinos with transient astrophysical phenomena \cite{IceCube:2018cha, Stein:2020xhk}.

In this work, we explore the possibility of using high-statistic neutrino events from a nearby Galactic supernova (SN) to probe for delayed neutrino signals induced by neutrino-DM interaction.
With upcoming detectors such as Hyper-Kamiokande and DUNE, the expected number of events should allow us to constrain non-standard interaction of neutrinos with low-mass DM via a new mediator. Here, we show that the time-delay induced by the DM-neutrino interaction would result in the late arrival of neutrinos between a day to a year after the first MeV neutrino burst is observed. This signature can be used to probe the interaction of neutrinos with DM particles in a mass range that is not easily accessible to other experiments.

\section{Method}\label{sec:method}
We consider a neutrino emitted by a source at a distance $D$, propagating through a bath of DM particles $\chi$. We define the optical depth $\tau=n_\chi\sigma_{\nu\chi} D$, where $n_\chi$ is the DM number density and $\sigma_{\nu\chi}$ is the total cross section for DM-neutrino interaction. Suppose that the interactions happen in the optically-thin limit, i.e., $\tau\ll 1$, such that neutrinos would at most experience one interaction as they travel towards the Earth. In this limit, if $N$ neutrinos are emitted at the source, the majority will arrive together, while a fraction of $\sim \tau N$ neutrinos will scatter and arrive later because of the increased trajectory length~\cite{Murase:2019xqi}. 
The time delay $t$ for the arrival of scattered neutrinos depends on the scattering angle, with a typical delay $\Delta t$ given by \cite{Murase:2019xqi}
\begin{equation}
    \Delta t \approx \frac{1}{2}\frac{\langle\theta^2\rangle}{4}D
    \simeq 1.3\times 10^7 {\rm s} \left(\frac{\langle\theta^2\rangle}{10^{-4}}\right)
    \left(\frac{D}{\rm 10 \, kpc}\right),
\end{equation}
where $\langle\theta^2\rangle$ is the mean of $\theta^2$, for a given differential cross section, and $\theta$ is the scattering angle. See also Refs.~\cite{echoana,echoMC}.

In the SN frame, DM is at rest and the incident neutrino's energy is $E_\nu$. For a scattering angle $\theta$, the scattered energy $E_\nu'$ is given by

\begin{equation}
    E_\nu'= \frac{E_\nu m_\chi}{m_\chi +E_\nu(1-\cos\theta)},
    \label{ScatteredEnergy}
\end{equation} 
where $m_\chi$ is the DM mass and we neglect neutrino mass. The differential cross section for a neutrino of incident energy $E_\nu$ to have a scattered energy $E_\nu^\prime$ is
\begin{equation}
\frac{d\sigma_{\nu\chi}}{dE_\nu^\prime}(E_\nu,E_\nu') = \frac{d\sigma_{\nu\chi}}{d\cos\theta}\frac{d\cos\theta}{dE_\nu^\prime},
\end{equation}
where
\begin{equation}
\frac{d\sigma_{\nu\chi}}{d\cos\theta}=\frac{1}{32\pi m_\chi^2}{\left(\frac{E_\nu'}{E_\nu}\right)}^2|{\mathcal M}|^2.
\end{equation}
Here the squared matrix element $|{\mathcal M}|^2$ depends on particle physics models that we discuss below. 
We will explore the range of mediator masses $\in [1\, {\rm eV},100\, {\rm MeV}]$ and $m_\chi\in [10\, {\rm eV},100\, {\rm keV}]$ in this work. 

We consider three particle physics models in this work. First, we consider fermionic DM that interacts via a vector mediator $V_\mu$, and the interaction Lagrangian of the form
\begin{equation}
\mathcal{L}_{\rm int} \supset g_\nu \bar\nu\gamma^\mu\nu V_\mu+
g_\chi \bar\chi\gamma^\mu\chi V_\mu,
\label{FermionDM_VectorMed_Lagrangian}
\end{equation}
where $g_\nu$ and $g_\chi$ are dimensionless coupling constants of the vector mediator to neutrinos and DM, respectively. Neutrino coupling to a vector mediator has also been used for example in the gauged $U(1)_{L_\mu-L_\tau}$ model \cite{Araki:2014ona,Escudero:2019gzq,Araki:2021xdk}. DM couplings to vector mediators have also been considered in the cosmological context \cite{Aarssen:2012fx,Cherry:2014xra,Duerr:2018mbd}. 
The Lagrangian in equation~\eqref{FermionDM_VectorMed_Lagrangian} implies that DM-neutrino scatterings are mainly forward scatterings, allowing us to remain within the small-angle scattering approximation. 
In addition, $m_V$ has little effect on the angular distribution for $m_V >5$~MeV. 

\begin{figure}
    \centering
    \includegraphics[width=0.7\textwidth]{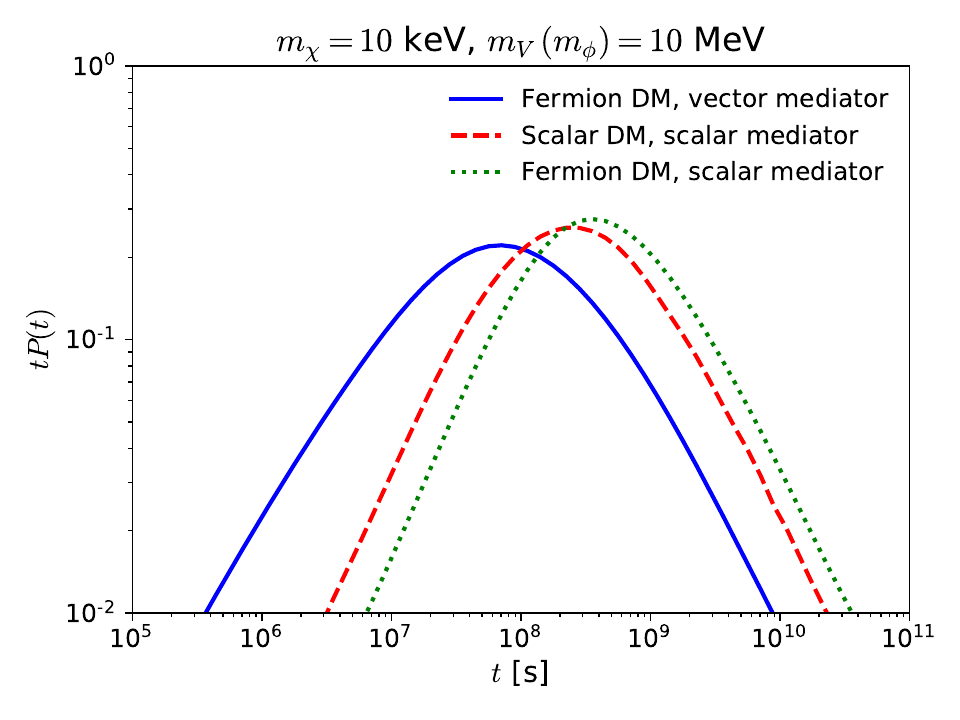}
    \caption{Time delay distribution of 15 MeV neutrinos for $m_\chi=10$~keV and a 10 MeV mediator. The distribution is multiplied by $t$. For each model, $g_\chi$ and $g_\nu$ have been chosen such that $\tau=10^{-3}$ for $D=10$ kpc. These results are, in fact, independent of $\tau$, provided that $\tau\ll 1$}
    \label{NuDMXSec}
\end{figure}

In addition to the Lagrangian in equation~\eqref{FermionDM_VectorMed_Lagrangian}, we will also consider fermionic DM with a scalar mediator
\begin{equation}
\mathcal{L}_{\rm int} \supset g_\nu \bar\nu \nu \phi+
g_\chi \bar\chi \chi \phi,
\label{FermionDM_ScalarMed_Lagrangian}
\end{equation}
and scalar DM with a scalar mediator
\begin{equation}
\mathcal{L}_{\rm int} \supset g_\nu \bar\nu \nu \phi+
g_\chi \Lambda\chi^* \chi\phi
\label{ScalarDM_ScalarMed_Lagrangian}.
\end{equation}
The differential and total cross sections for these interactions can be found in Ref.~\cite{Arguelles:2017atb}. For the last Lagrangian, we note that the coupling is split into a dimensionless coupling $g_\chi$ and an energy scale $\Lambda=100$~GeV. 
In scalar mediator models, if neutrinos are Dirac fermions we would need to consider mixing with sterile states; for Majorana neutrinos, $\bar{\nu}\nu$ should be interpreted as $\overline{\nu}_L^c\nu_L/2 + {\rm c.c}$. 
For example, DM interactions with scalar mediators arise in Standard Model extensions, where DM couples to the Higgs sector and protected by a $\mathbb{Z}_2$ symmetry \cite{Arcadi:2021mag}. This leads to Higgs portal models for fermionic DM \cite{Kouvaris:2014uoa} or scalar DM \cite{Holz:2001cb,Arcadi:2021mag} after the spontaneous electroweak symmetry breaking.

Let $P(t)$ be the probability density function of the neutrino time delay $t$ of the neutrinos within the arrival energy range of interest. By working in the $\tau\ll 1$ limit, $P(t)$ depends on $(1/\sigma_{\nu\chi})d\sigma_{\nu\chi}/d\cos\theta$ and is therefore independent of the coupling strength \cite{Hatchett:1978,Murase:2019xqi,echoana}.
In Figure~\ref{NuDMXSec} we show the time delay distribution of neutrinos with an initial energy of 15 MeV, a dark matter mass $m_\chi=10$ keV and a 10 MeV mediator. Each model has values of $g_\chi$ and $g_\nu$ such that $\tau = 10^{-3}$, in the optically-thin regime. Here we see that among the three models, the fermionic DM with a vector (scalar) mediator yields shorter (longer) time delays. This is related to the details of the angular distribution of the model, where smaller scattering angles lead to shorter time delays. For fermionic DM with a scalar mediator, we see that the $\Delta T$ decreases for DM masses above 100 keV and $m_\phi$ in the 100 eV -- 10 keV range. This decrease is caused by the energy threshold used in the analysis, which we address in the next section.

In order to estimate the temporal profile for the arrival of neutrinos from a SN, we adopt a SN neutrino spectrum at the source of the form \cite{Keil:2002in,Mirizzi:2015eza,LunardiniSNReview},
\begin{equation}
\Phi_{\nu}(E_\nu) = \frac{L_\nu}{\langle E_\nu\rangle^2}\frac{(\alpha+1)^{\alpha+1}}{ \Gamma(\alpha+1)}\left(\frac{E_\nu}{\langle E_\nu\rangle}\right)^\alpha
\exp\left(-\frac{(\alpha+1)E_\nu}{\langle E_\nu\rangle}\right),
\label{SN_NuSpectrum}
\end{equation}
where $\langle E_\nu\rangle$ is the average neutrino energy, $\alpha$ is a pinching parameter, $L_{\bar{\nu}_\alpha}$ is the neutrino luminosity, and $\Gamma$ is the Euler Gamma function. Hereafter, we assume $\alpha=2.3$ and $\langle E_\nu\rangle =$ 16 MeV, although in general the values of $\alpha$ and $\langle E_\nu\rangle$ are different among neutrino flavors \cite{Keil:2002in}. The total neutrino energy is set to $\mathcal{E}_\nu = L_\nu T_{\rm dur}=3\times 10^{53}$ erg, where $T_{\rm dur}=10$~s is the duration of the neutrino emission.Neutrino emission consists of several stages. Around the core bounce, the so-called $\nu_e$ neutronization burst is expected, which lasts for $\sim20$~ms. This is followed by the accretion phase with significant production of $\nu_e$ and $\bar{\nu}_e$, which lasts for a few seconds (e.g., \cite{Nakazato:2012qf,Tamborra:2014hga,Blum:2016afe}). Then, the protoneutron star cools and neutrino luminosities of all flavors become similar, lasting for $\sim10-100$~s (e.g., \cite{Perego2015,Mirizzi:2015eza,Suwa:2020nee}). The total energy we are considering here can also be matched to the simulation results presented in Ref.~\cite{HK:2021frf} within 1~s after the bounce.

The supernova spectrum consists of $\bar{\nu}_e$ and $\bar{\nu}_x$ (non-electron antineutrinos). 
We assume that both fluxes are related by $\Phi_{\bar{\nu}_x} = 0.3 \Phi_{\bar{\nu}_e}$ \cite{Blum:2016afe}, such that they have the same production spectra. This assumption is made for simplicity because using separate spectra would require us to look at $\bar{\nu}_e$ and $\bar{\nu}_x$ with different pinching parameters $\alpha$. The flux is normalized so the total neutrino energy in all three flavors is equal to $\mathcal{E}_\nu$.
For pure adiabatic transitions, the fluxes at the surface of the star are $\Phi_{\bar{\nu}_1}=\Phi_{\bar{\nu}_e}$ and $\Phi_{\bar{\nu}_2}=\Phi_{\bar{\nu}_3} = \Phi_{\bar{\nu}_x}$, assuming normal mass ordering \cite{Dighe:1999bi}. The $\bar{\nu}_e$ flux on Earth becomes $\Phi_{\bar{\nu}_e} = \sum_i \Phi_{\bar{\nu}_i} |U_{ei}|^2$, where $U$ is the neutrino mixing matrix.

For a nearby SN of $D\sim\mathcal{O}(10)$ kpc, we can assume a local DM density $n_\chi=0.3$ cm$^{-3}(m_\chi/1 \, {\rm GeV})^{-1}$. As we show in our results, within our parameter space the typical time delays would lie in the $10^2$-$10^8$~s range.
While the SN neutrino spectrum is time-dependent (see e.g., Ref.~\cite{HK:2021frf}), the characteristic time delays are much longer than $T_{\rm dur}$, so we use the time-integrated flux on Earth
\begin{equation}
    \frac{dN_\nu}{dE_\nu} = \frac{\Phi T_{\rm dur}}{4\pi D^2}.
\end{equation}
The number of neutrino events in Hyper-Kamiokande is 
\begin{equation}
N_{\rm events} = N_{\rm T}\int_{10 \, \rm MeV}^{50 \, \rm MeV} \frac{dN_{\bar{\nu}_e}}{dE_{\bar{\nu}_e}}\sigma_{\rm QE}(E_\nu)dE_{\bar{\nu}_e},
\label{NEvts}
\end{equation}
where $N_{\rm T}$ is the number of targets ($1.25\times 10^{34}$ for the 187 kton HK detector fiducial volume \cite{HK:2021frf}) and $\sigma_{\rm QE}$ is the quasi-elastic inverse beta decay cross section. We have assumed 14 MeV as the neutrino energy threshold. These are the total number of events, which accounts for both scattered and unscattered neutrinos. For our chosen parameter set, we get $N_{\rm events}=48200$. This is consistent with Ref.~\cite{HK:2021frf}, which obtained $N_{\rm events} \approx 20000$ for $T_{\rm dur}=500$~ms, although we use a larger total neutrino energy (in all flavors) of $\mathcal{E}_\nu = 3\times 10^{53}$~erg. 

To calculate the delayed neutrino spectrum $dN_{\rm scatt}/dE_\nu$, which is the time-integrated spectrum of all scattered neutrinos, we use
\begin{equation}
    \frac{dN_{\rm scatt}}{dE_\nu} = \int_0^\infty dt \int_{E_{\nu}}^{E_\nu^{\prime {\rm max}}(E_\nu,t)} dE_\nu^\prime \frac{dN_\nu}{dE_\nu^\prime} P(t,E'_\nu)  \frac{d\sigma_{\nu \chi}}{dE_\nu}(E_\nu^\prime,E_\nu) n_\chi D,
\end{equation}
where the integrand is the product of the SN spectrum at $E_\nu^\prime$ and the probability that a neutrino of energy $E_\nu^\prime$ scatters once and arrives with an energy $E_\nu$, in the optically-thin limit. The spectrum $dN_\nu/dE_\nu$ is inserted into equation~\eqref{NEvts} to obtain the number of events that undergo scatterings.
We point out that the relationship between $N_{\rm scatt}$ and $N_{\rm events}$ is not trivial due to the threshold, as some of the scattered neutrinos will fall below that energy,  but the relationship $N_{\rm scatt}\sim\tau N_{\rm events}$ provides an order of magnitude estimate. We may write $N_{\rm scatt} = \kappa \tau N_{\rm events}$, where $\kappa$ is the fraction of scattered events with $E_\nu>$ 14 MeV and is determined from simulations. Given that effects of 
$P(t,E_\nu)$ and $E_\nu^{\prime \rm max}$ are included in $\kappa$, the rest will only depend on $m_\chi$ and the mediator mass when $\tau\ll 1$. Within this approximation, for fixed DM and mediator masses, we have $N_{\rm scatt}\propto g_\nu^2g_\chi^2 L_\nu T_{\rm dur}/D$.

We constrain the $(g,m_V,m_\chi)$ parameter space under the assumption that no significant background excess has been observed within a time window $\Delta T$ after the SN neutrino burst is detected. 

\begin{figure}
    \centering
    \includegraphics[width=0.8\textwidth]{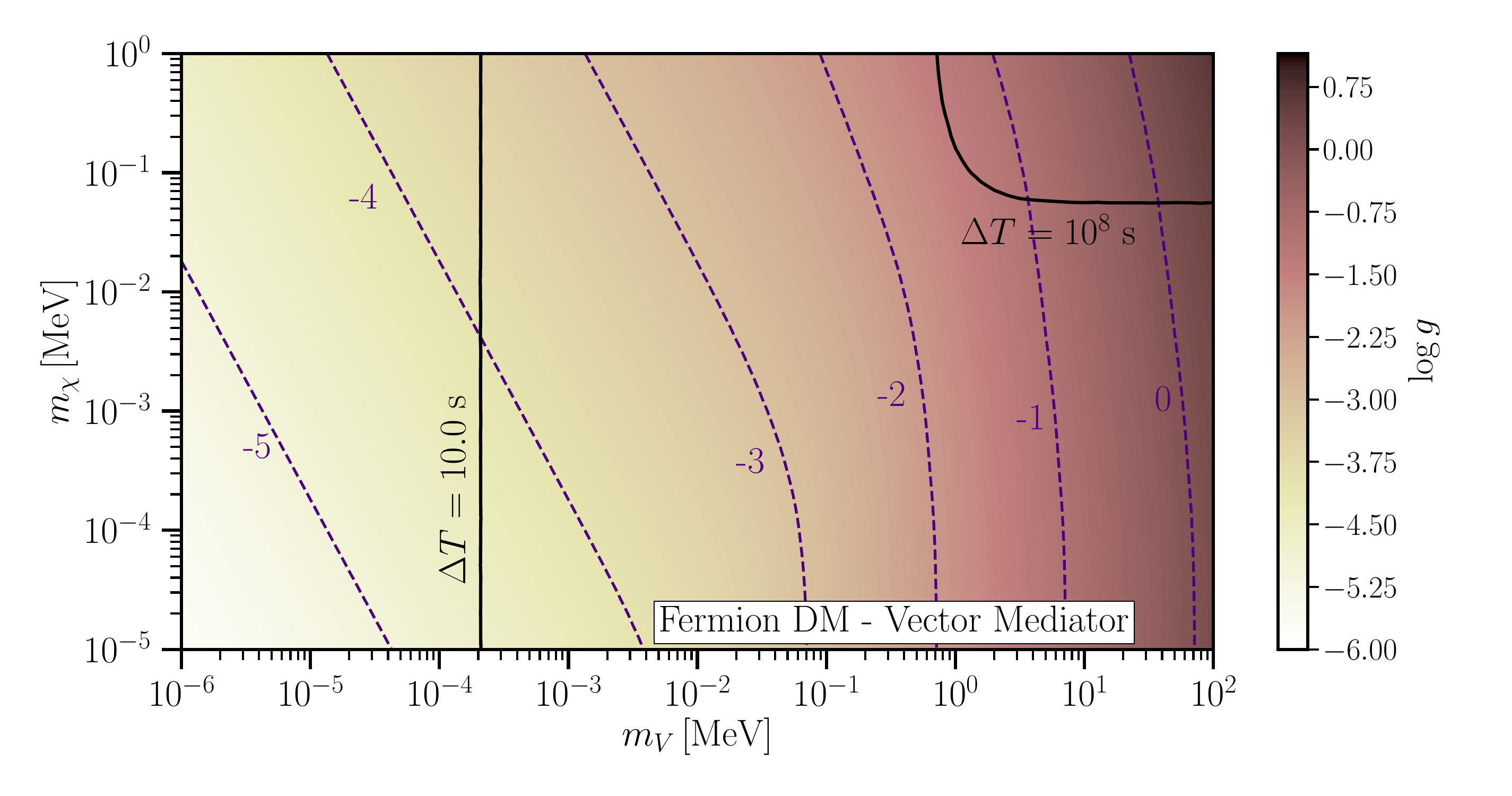}
    \includegraphics[width=0.8\textwidth]{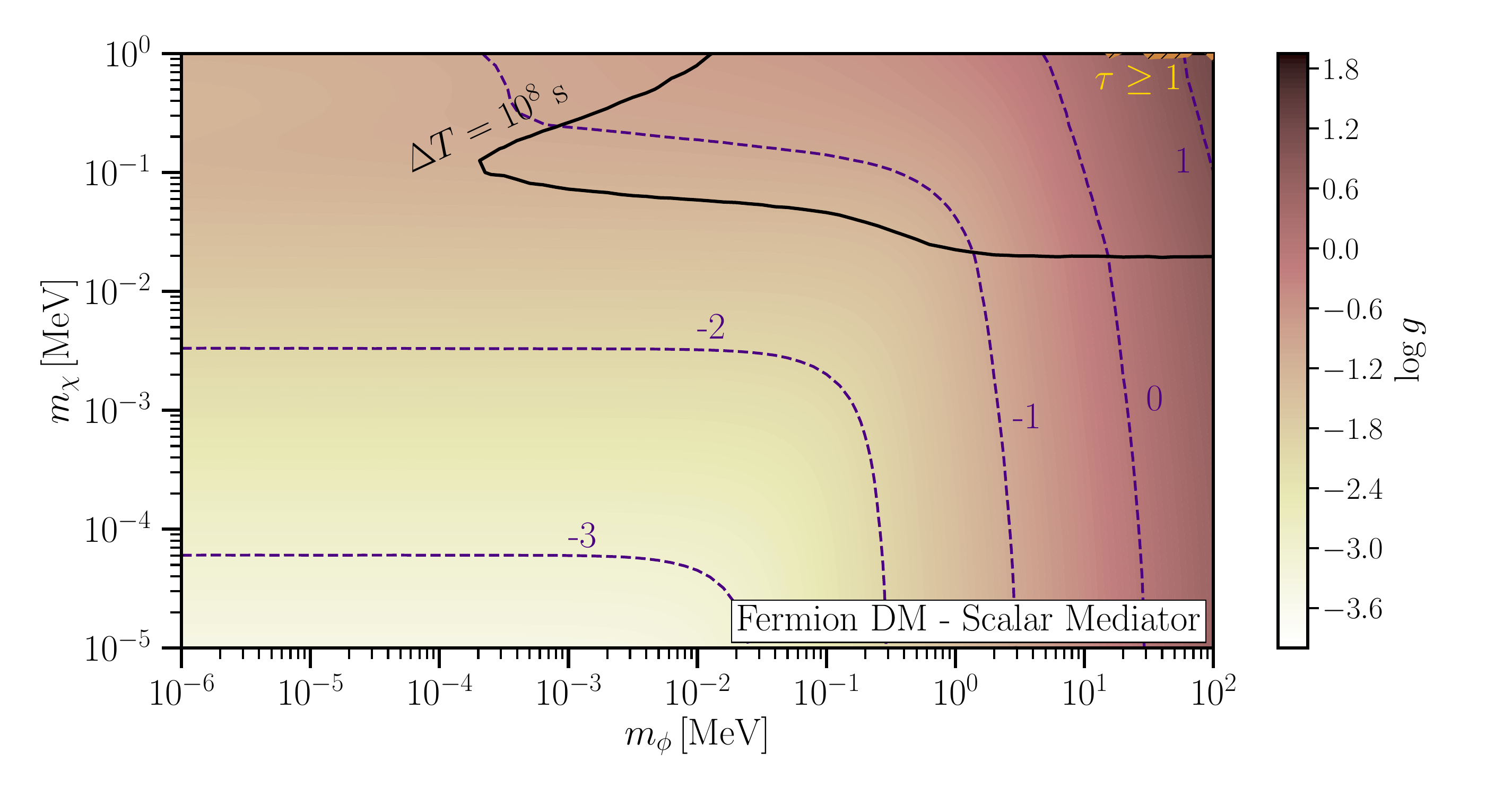}
    \includegraphics[width=0.8\textwidth]{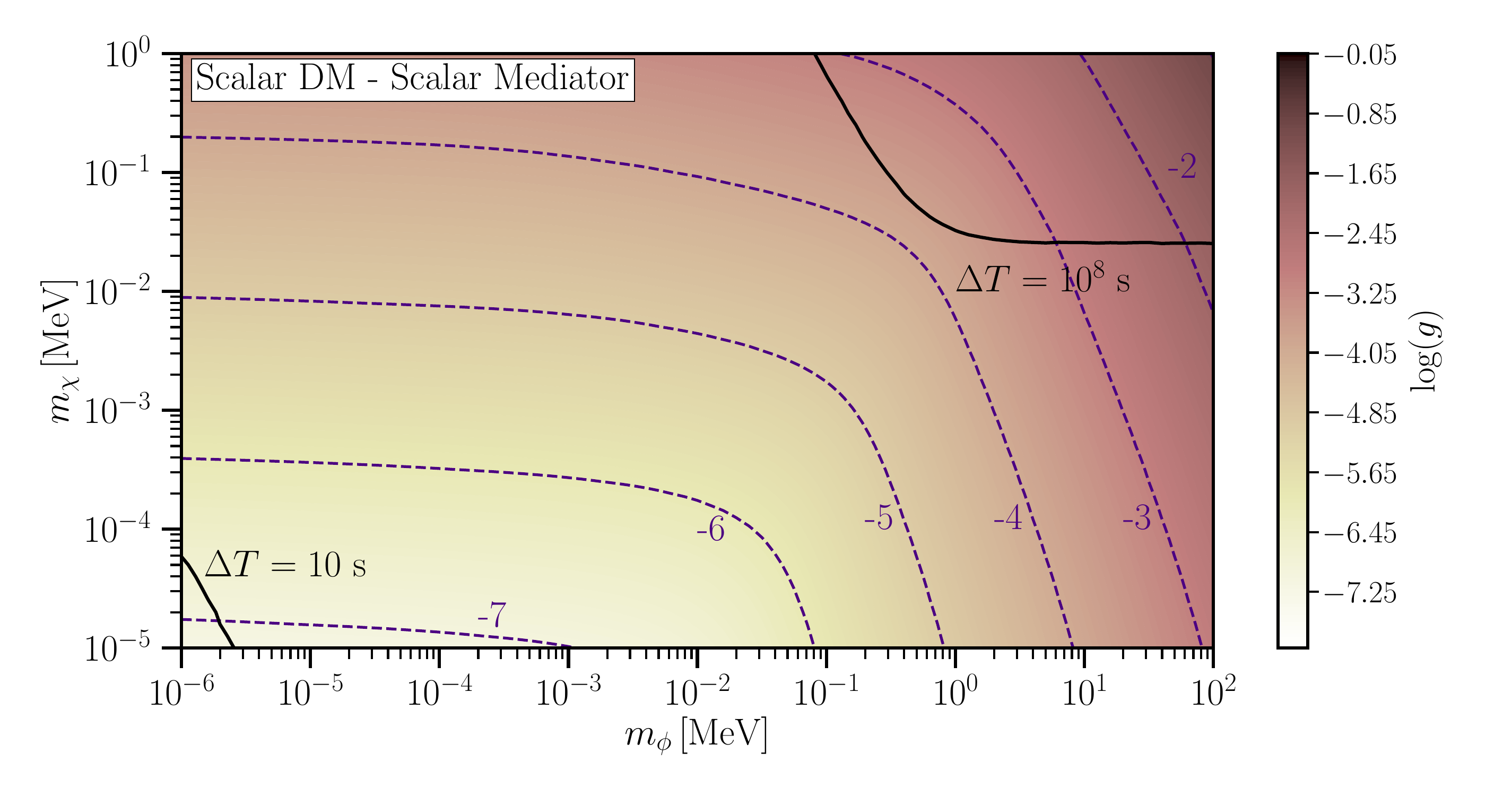}
    \caption{Neutrino-DM coupling constraints on $g=(g_\nu g_\chi)^{1/2}$ for the models described by equations~\eqref{FermionDM_VectorMed_Lagrangian} (top), \eqref{FermionDM_ScalarMed_Lagrangian} (middle) and \eqref{ScalarDM_ScalarMed_Lagrangian} (bottom). The time window $\Delta T$ is the time taken to enclose $50\%$ of the scattered neutrinos with energy above 14 MeV. The region $\tau\geq 1$ has been shaded for the Fermion DM and Scalar Mediator case. The other models do not have $\tau\geq 1$ within the parameter space shown.}
    \label{NuDMConstraints}
    
\end{figure}

We use the Feldman-Cousins upper limits \cite{Feldman:1997qc} to obtain constraints on the parameters of DM-neutrino interactions. The background rate is obtained from the different channels provided in \cite{Lin:2019piz}: invisible muons, neutral current, atmospheric neutrinos, lithium, reactor neutrinos and diffuse supernova neutrinos. For a 187 kton detector with Gadolinium in the energy range [14 MeV, 50 MeV], the total rate would be 3.41$\times 10^{-6}$ Hz. We use this rate to estimate the expected number of background events $\mu_b$ over a time $\Delta T$ after the MeV burst. Starting from the arrival time of the unscattered signal, we take the time window $\Delta T$ that encloses a factor $0<\beta\leq 1$ of $N_{\rm scatt}$. Our calculations of $\Delta T$ enforce an $E_\nu>14$ MeV threshold. The different interaction models affect $\Delta T$ only through the distribution $P(t)$, so the choice of time window depends on DM and mediator masses only. Within this $\Delta T$, we find the Feldman-Cousins upper limit $\mu_s$, assuming that the expected number of events is $\mu_b$ (i.e., background only) and the observed number of events is also $\mu_b$ . This $\mu_s$ would then correspond to the expected number of delayed neutrino events within $\Delta T$. We set $\mu_s=\beta N_{\rm scatt}$ and adjust $g_\nu^2g_\chi^2$ to get this equality to hold. This equation for $\mu_s$ relies on the scaling of $N_{\rm{scatt}}\propto g_\nu^2g_\chi^2$, which is only valid in the optically thin regime. Therefore, this method cannot be applied for $\tau\geq 1$, which correspond to the shaded regions in Figure \ref{NuDMConstraints}. In the case of heavier dark matter $m_\chi\gtrsim 100$~keV, $\Delta T>10^8$ s for heavy mediators. For these cases, we set $\Delta T=10^8$ s and adjust $\beta$ accordingly.  

\section{Results}
We calculate the $95\%$ confidence level (CL) upper limit on the coupling constant in Hyper-Kamiokande in the event of a 10~kpc SN, assuming no significant delayed neutrino signal is detected. For this purpose, we define the effective coupling $g=(g_\nu g_\chi)^{1/2}$, and set $\beta=0.5$. Our choice of $\beta$ is motivated by the Monte Carlo simulation results \cite{echoMC} which suggest that $\Delta T$ will coincide with the peak of the $t P(t)$ distribution (see figure~\ref{NuDMXSec} as an example). 
We show the upper limit on the coupling for different DM and mediator mass in figure~\ref{NuDMConstraints}. 
We mark the set of points for which $\Delta T=10^8$~s and $\Delta T = 10$~s, where the latter corresponds to the duration of the neutrino emission. We also shade the region where $\tau\geq 1$, where the optically thin approximation is not satisfied and our results are not applicable. This region is only present within our parameter space when we assume fermionic DM with a scalar mediator. For fermionic DM and a vector mediator, we see that when $m_{\rm V}\lesssim 100\,{\rm eV}$, the time window used is too short for the delayed signal to be well separated from the initial MeV burst. This is not the case for scalar mediators, where the scattering angles remain relatively large for very light scalars. 

We also compare our constraints against limits from other observables. The first one is the neutrino self-interaction bound $g_{\tau\tau}<0.27$ from \cite{Blinov:2019gcj}, which only applies to $\nu_\tau$ coupling. To convert $g_{\tau\tau}$ into an effective coupling $g_\nu^{\rm eff}$ and then into the $\nu$-DM coupling $g$, we proceed as follows. The $\bar{\nu}_e$ flux from $\nu-$DM scatterings, $\Phi_{\bar{\nu}_e,{\rm scatt}}$, is given by the probability that $\bar{\nu}_i$ interacts via $g_{\tau\tau}$, becomes $\bar{\nu}_j$ and is detected on Earth as $\bar{\nu}_e$. In the limit $\tau\ll 1$, this probability is simply an effective optical depth. We then write $\Phi_{\bar{\nu}_e,{\rm scatt}}=\sum_i \tau_i \Phi_{\bar{\nu}_i}$, which is the sum of fluxes of scattered $\bar{\nu}_i$ which are detected as $\bar{\nu}_e$. With the assumption that $\bar{\nu}_e$ and $\bar{\nu}_x$ are proportional to each other, we may also simplify this expression to $\Phi_{\bar{\nu}_e,{\rm scatt}}=\tau^{\rm eff}\Phi_{\bar{\nu}_e}$ for an effective optical depth 

\begin{equation}
\tau^{\rm eff} = n_\chi\sigma_{\nu\chi}^{\rm eff}D = n_\chi\frac{\sigma_{\nu\chi}D}{g_\nu^2}\sum_{i,j} |U_{ej}|^2 |U_{\tau j}|^2 |U_{\tau i}|^2 g_{\tau\tau}^2 P_i,
\end{equation}
where $P_i = \Phi_{\bar{\nu}_i}/\Phi_{\bar{\nu}_e}$. Note that the cross sections involved have negligible contributions from neutrino mass, so $\sigma_{\nu\chi}$ is the same regardless of the neutrino mass eigenstate involved. The cross section $\sigma_{\nu\chi}^{\rm eff}$ is now proportional to $(g_\nu^{\rm eff})^2$. We can thus absorb neutrino mixing effects into this coupling, such that
\begin{equation}
(g_\nu^{\rm eff})^2 = g_{\tau\tau}^2 \sum_{i,j} |U_{ej}|^2 |U_{\tau j}|^2 |U_{\tau i}|^2 P_i.
\end{equation}
With the current values of the oscillation parameters, this leads us to $g_\nu^{\rm eff} = 0.1$. To get the upper bound of $g$, we use $g_\nu^{\rm eff}$ together with the upper bound $g_\chi< 4\pi$ originating from the perturbative limit. This then leads to a bound $g<\sqrt{4 \pi g_{\nu}^{\rm eff}}=1.11$.

We also have the BBN constraint on mediator masses, given in\cite{Escudero:2019gzq,Blinov:2019gcj}. Finally, we also consider the constraints for merging galaxy clusters, which requires  $\sigma_{\chi\chi}/m_\chi<0.1$ cm$^2$ g$^{-1}$ \cite{PhysRevD.87.115007,PhysRevLett.110.111301}. Here $\sigma_{\chi\chi}$ is DM self-scattering cross section in the low velocity limit. The cluster constraints provide upper bounds on $g_\chi$ for fixed DM and mediator masses. To convert this into a bound for $g$, we  need to assume a ratio $g_\nu/g_\chi$. Since the constraints are on $g_\chi$ only, choosing small (large) $g_\nu/g_\chi$ will strengthen (weaken) the bounds on $g$.

\begin{figure*}
    \centerline{
    \includegraphics[width=0.5\columnwidth]{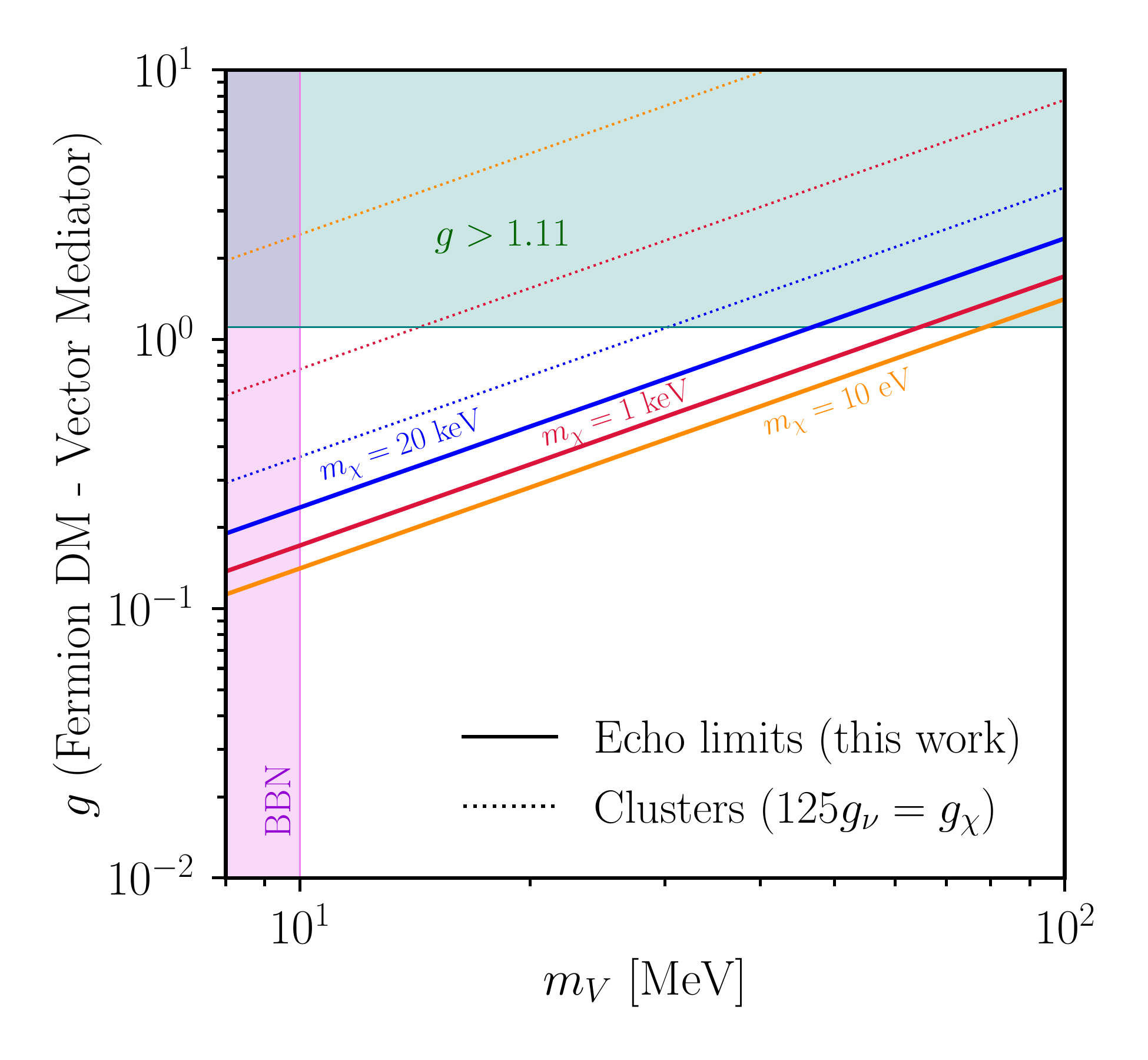}
    \includegraphics[width=0.5\columnwidth]{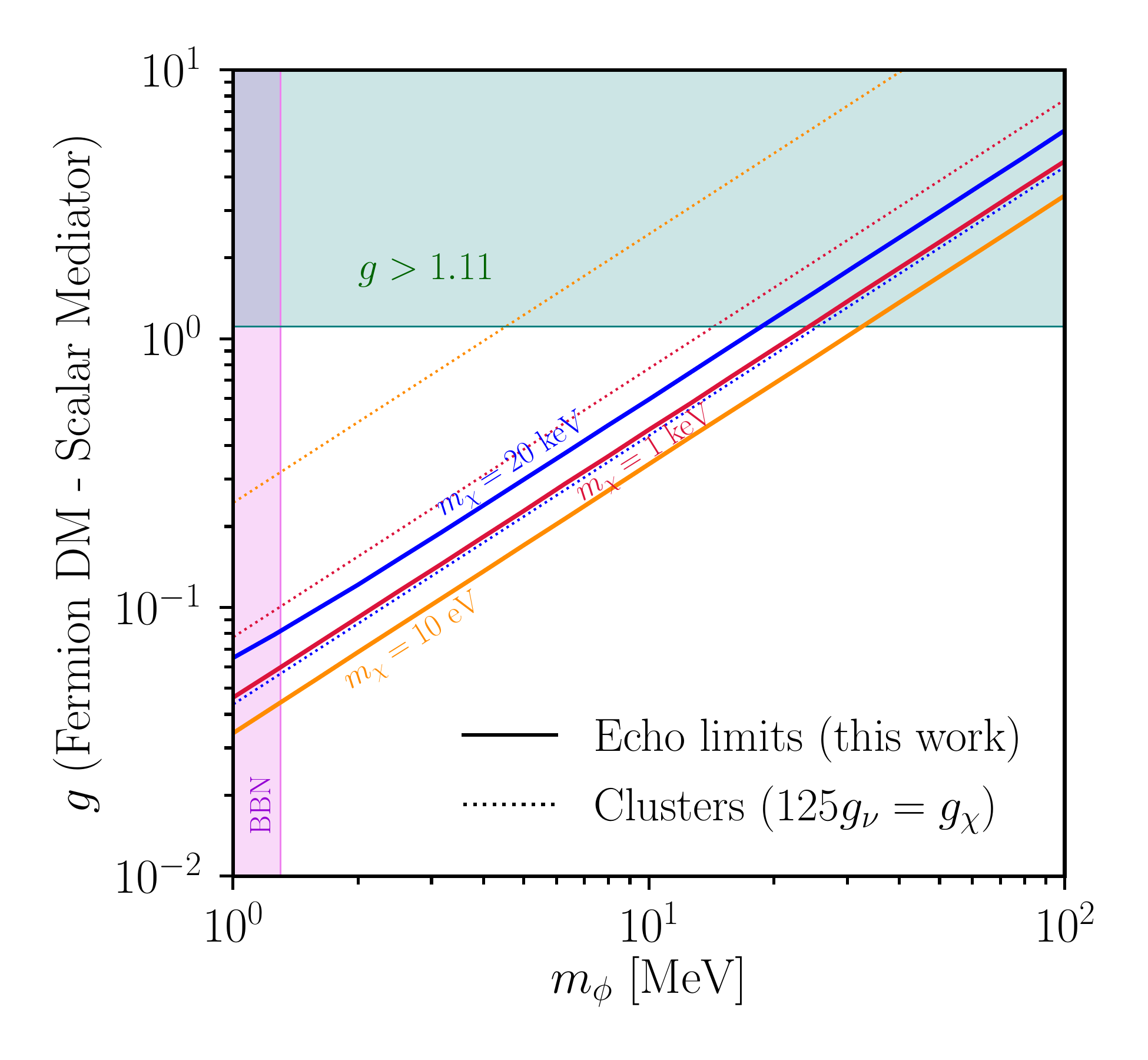}
    }
    \caption{Neutrino-DM coupling constraints on $g=(g_\nu g_\chi)^{1/2}$ for the models described by equations~\eqref{FermionDM_VectorMed_Lagrangian} (left panel) and \eqref{FermionDM_ScalarMed_Lagrangian} (right panel). Cluster constraints are shown as dotted lines for each DM mass, assuming $125 g_\nu=g_\chi$, corresponding to the ratio of $g_\nu = 0.1$ and $g_\chi = 4\pi$. BBN nucleosynthesis constraints \cite{Escudero:2019gzq,Blinov:2019gcj} correspond to the shaded region (magenta).}
    \label{NuDMConstraints_2}
    
\end{figure*}

\begin{figure*}
    \centerline{
    \includegraphics[width=0.7\columnwidth]{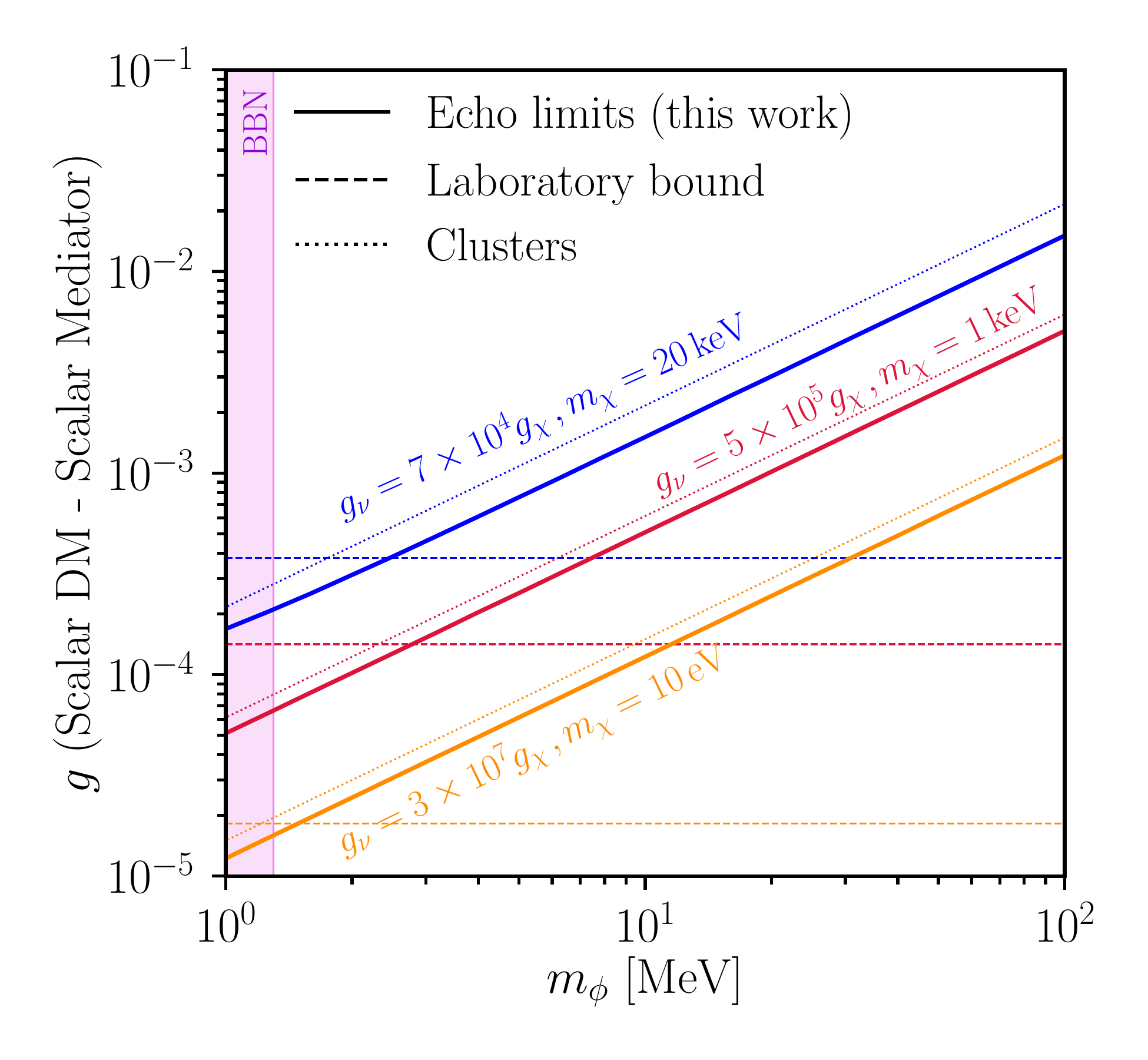}
    }
    \caption{Neutrino-DM coupling constraints on $g=(g_\nu g_\chi)^{1/2}$ for scalar DM and scalar mediator. Here, we present the constraints for three DM masses. BBN constraints \cite{Escudero:2019gzq,Blinov:2019gcj} correspond to the shaded region (magenta). The cluster lines (dotted) show the limits for different ratios of $g_\nu$ and $g_\chi$. The laboratory bounds with the same ratios used in each cluster line are shown as dotted lines. The energy scale is $\Lambda=100$~GeV.}
    \label{NuDMConstraints_3}
\end{figure*}

In figure~\ref{NuDMConstraints_2}, we show the 2D projections for selected DM masses assuming fermionic DM and show the aforementioned bounds from laboratory measurements, BBN and cluster constraints.
For the case of fermionic DM, we take $g_\nu/g_\chi = 1/125$, the ratio corresponding to $g_\chi = 4\pi$ and $g_\nu = 0.1$. For this choice, our constraints are stronger than laboratory and cluster bounds for $m_\chi<20$ keV ($m_\chi<1$ keV) for a vector (scalar) mediator. In the case of scalar DM with a scalar mediator, as shown in figure \ref{NuDMConstraints_3}, we find that the cluster bounds required a $g_\nu/g_\chi$ ratio above $10^4$ for the echo limits to be competitive. For these large ratios, we easily reach the laboratory bound on $g_\nu$, as shown by the dashed lines. In the end, our constraints are stronger than laboratory and cluster bounds when $m_\phi\lesssim$ 3 MeV for $g_\nu/g_\chi=7\times 10^4$($g_\nu/g_\chi=5\times 10^5$) for $m_\chi = 20 (1) $ keV. For $m_\chi=10$ eV and $g_\nu/g_\chi = 3\times 10^7$, laboratory bounds are stronger than our bounds when $m_\phi\gtrsim$ 1.3 MeV. Overall, given the BBN bound, there is a limited range of $m_\phi$, in which the constraints in our work are the strongest.

If the water detector does not have Gadolinium, the energy threshold would be at around 17 MeV, since below it the spallation background is large \cite{Lin:2019piz}. Taking this energy threshold into account, the background is dominated by invisible muons, increasing $\mu_b$. Likewise, the increased energy threshold means that a larger fraction of the scattered events will lie below it. In this scenario, depending on $m_\chi$ and mediator mass, we may require up to 3 times as many scattered events. In turn, the couplings presented in our results would have to be increased by up to $30\%$.

A general feature is that for a fixed $m_\chi$, the constraint on the coupling weakens for larger mediator masses. The delayed neutrino spectrum is proportional to $\sigma_{\nu\chi}$ in the small optical depth limit, so a larger value of $g$ is needed to account for heavier mediators. 
On the other hand, for a fixed mediator mass, the constraint gets weaker for heavier DM and this weakening becomes more dramatic for lighter mediators. The total cross section monotonically decreases with $m_\chi$, which contributes to weaker constraints. In all three models considered, we see that there is a region of parameter space that is not constrained by BBN or laboratory measurements that can be probed by the echo approach. 

\begin{figure}
    \centering
    \includegraphics[width=0.7\columnwidth]{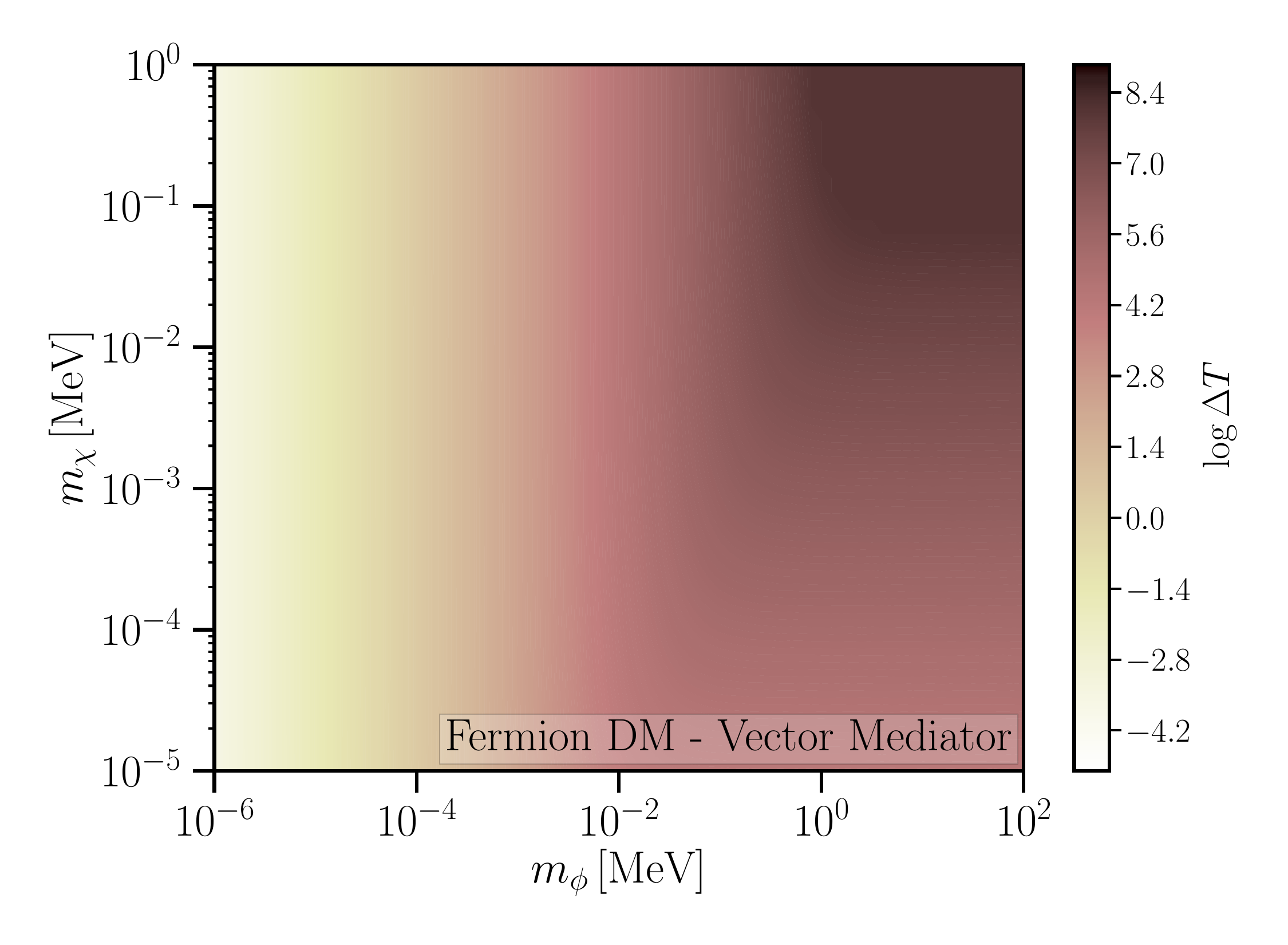}
    \caption{Time window $\Delta T$ as a function of the mediator mass and DM mass, for 50\% of the delayed neutrino signal to reach Earth in the event of a SN at a distance of 10 kpc. We show the case for fermion DM and a vector mediator.}
    \label{DeltaTConstraints}
\end{figure}

The time window $\Delta T$ used to constrain $g$ is shown in figure~\ref{DeltaTConstraints} for fermionic DM with a vector mediator. We find that for $\mathcal{O}(10$ keV) mediators, we need time delays between a few weeks and a month. We see that $\Delta T$ goes up to a year for mediators heavier than 100 keV, and remains constant for a fixed $m_\chi$, for which the angular distribution becomes less dependent on $m_V$. The typical scattering angle is sensitive to $m_\chi$, and heavier DM monotonically increases the time window $\Delta T$ to achieve a given $\beta$, for a fixed mediator mass. For DM above 100 keV, however, a local maximum can be reached and then $\Delta T$ decreases. The reason behind this is that $\Delta T$ is determined by the delayed neutrino signal, which incorporates a neutrino energy threshold of 14 MeV. As $m_\chi$ increases, the scattering angle increases and a larger number of neutrinos are scattered to energies below the threshold. Neutrinos below the energy threshold are not considered part of the delay distribution used to determine $\Delta T$, and the removal of these events with large delays causes $\Delta T$ to decrease. This effect is clearly visible in the case of fermionic DM with a scalar mediator (see figure~\ref{NuDMConstraints}), where the time delay distributions tend to have a large peak close to $10^8$ s for $m_\chi>100$ keV, even for light scalar masses.

\begin{figure}
    \centering
    \includegraphics[width=0.6\textwidth]{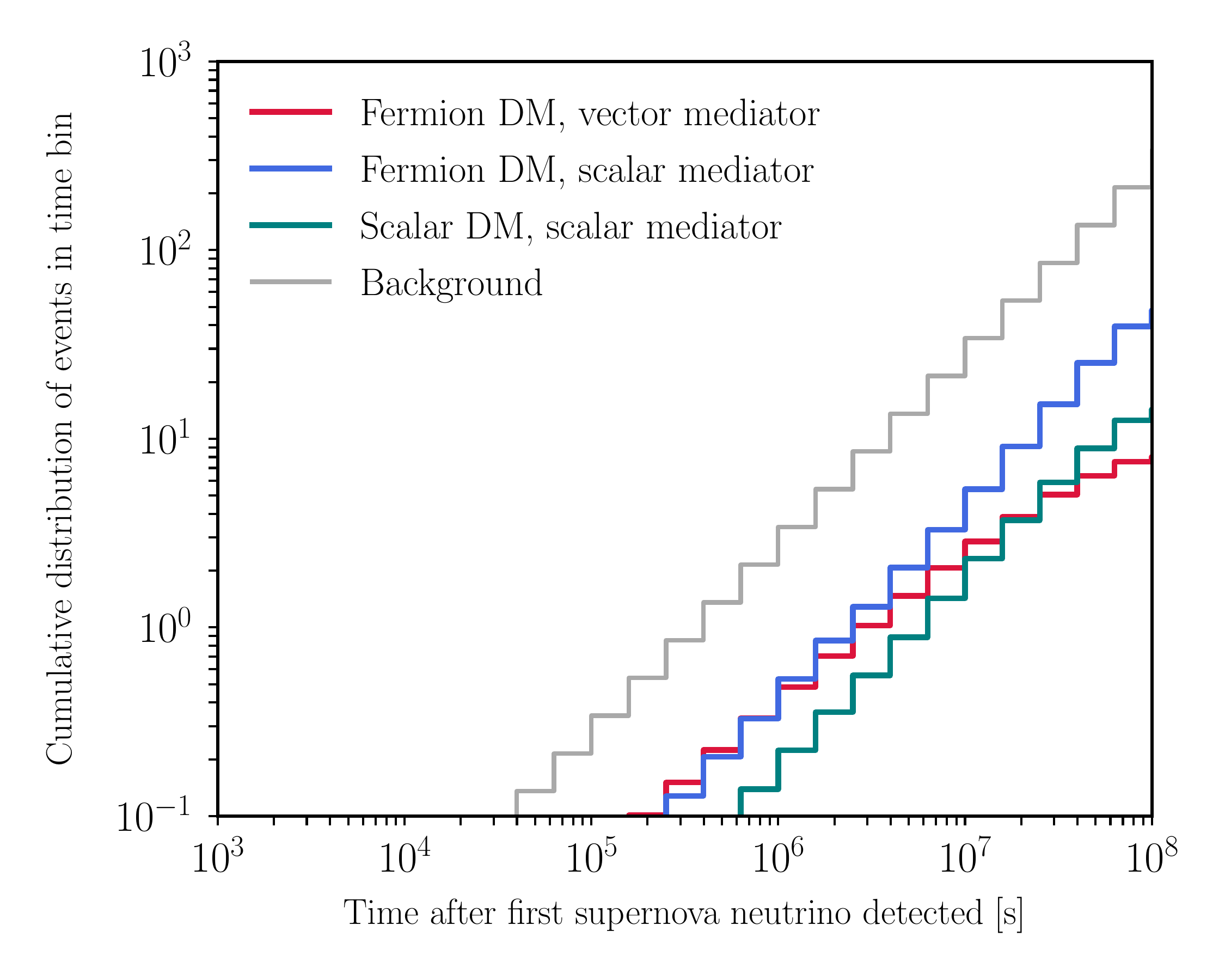}
    \caption{Cumulative number of delayed signal events in Hyper-Kamiokande, compared to the background, which has a rate of 3.41$\times 10^{-6}$~Hz. Here, we show the three different models used for $m_\chi=10\, {\rm keV}, m_{\rm V} = 10\, {\rm MeV}, m_{\phi} = 10\, {\rm MeV}$. Each distribution has a total number of events corresponding to the time windows used for our upper limits. In the case of the vector mediator, this corresponds to 10 events and $\Delta T = 1.7\times 10^6$ s; for the scalar mediator with scalar (fermionic) DM, this is  17 (58) events and $\Delta T = 3.9 \times 10^6$ s ($5.1\times 10^6$ s). 
    }
    \label{EventRates}
\end{figure}

To get a better view of the comparison between signal and background events, we present in figure~\ref{EventRates} the cumulative number of signal events for 10 keV DM and a 10 MeV mediator. This choice of these parameters corresponds to the region with longer time delays, where delays get closer to $10^8$~s. Each model is normalized to the number of events required by the Feldman-Cousins upper limit. Similar to figure~\ref{NuDMXSec}, the vector mediator has several neutrino events early on, as the forward scattering is predominant. Even though the signal to background ratio is quite small, the upper limit $\mu_s$ grows roughly with $\sqrt{\mu_b}$, so fewer signal events are needed. 

We note that the constraints shown were obtained for a SN with total neutrino energy of $3\times 10^{53}$ erg at $D=10$ kpc. Our approach relies on determining $\mu_s$, which depends only on the chosen time window (i.e. on $m_\chi$ and $m_V$. Once $m_\chi$ and $m_V$ are fixed, $N_{\rm scatt} \propto g_\nu^2g_\chi^2 \mathcal{E_\nu}/D$, so we can get constraints for other SNe by the appropriate scaling. Thus, choosing different SNe models, namely changing $\alpha$ and $\langle E_\nu\rangle$, mildly affects the constraints, as long as the majority of the SN neutrinos is above the neutrino energy threshold for Hyper-Kamiokande.

\section{Discussion}
The presence of DM-neutrino interaction may affect the effective number of relativistic species, $N_{\rm eff}$, which provides additional constraints.  
If DM particles are in equilibrium with the SM bath prior to the neutrino-photon decoupling, the dark matter mass is constrained to be $m_{\chi}>1$~MeV \cite{Blinov:2019gcj}. However, it was shown that if the equilibrium between DM and the SM neutrinos occurs after the neutrino-photon decoupling, then $N_{\rm eff}$ constraints on the interactions can be significantly relaxed \cite{Berlin:2017ftj}, allowing for sub-MeV DM.

We also point out that, contrary to the assumption in \cite{Blinov:2019gcj}, the parameter space for the models presented also covers the region where DM is lighter than the mediator, in which case DM freeze-out through $\chi-\chi$ annihilation to two mediators is kinematically forbidden. Also, as we provide bounds on the effective coupling $g$, $g_\chi$ could be a lower value to be compliant with other constraints by increasing $g_\nu$ or vice versa. Note that, due to this interplay between both couplings, it is possible for our constraints to provide competitive or stronger bounds than clusters constraints.

 If we relax the assumption of a homogenous DM density, we would need to perform a column integral of $n_\chi\sigma_{\nu\chi}$ to get the optical depth. In this scenario, the neutrino is more likely to interact in the regions with the largest DM density.  In particular, if the source is located such that the signal has to cross the Galactic Center, the optical depth would increase by a factor of $\sim 20$ compared to the assumption of constant DM density \cite{Murase:2012xs}. Since our number of scattered events is proportional to $\tau$, we would expect our constraints on the coupling to be stronger by a factor $\sim 20^{1/4}\approx 2$. When $m_\phi>1$ MeV and $m_\chi>$100 keV and the delays become larger than $10^8$ s, the increased optical depth may not give a stronger constraint. The time delay also depends on where the scattering takes place. If a very dense DM region is located close to the source such that the scattering is likely to occur far away from Earth, the typical time delay will be longer. For heavier DM, longer delays would force us to adopt $\Delta T = 10^8$ s and would begin to lose signal events to this time cut, which in turn can weaken our constraints.

For the specific case of SN 1987A, neutrino-DM interaction constraints are discussed in Ref.~\cite{Mangano:2006mp}. 
For MeV DM, it was found that for a constant scattering cross section, cosmological data provide stronger bounds than SN1987A data. As the total number of neutrinos detected from this SN is relatively small, the bounds are obtained from the assumption that there was no significant neutrino absorption in the observed spectrum. Compared to the bound on the cross section $\sigma_{\nu\chi}/m_\chi<10^{-25}$ cm$^2$ MeV$^{-1}$ from SN1987A, our projected bound with HK is  $\sigma_{\nu\chi}/m_\chi<1.2\times 10^{-27}$ cm$^2$ MeV$^{-1}$ for fermionic DM and a scalar mediator case, with $m_\chi = 1$ keV and $m_\phi = 10$ MeV. For this projected bound we assumed a neutrino energy $E_\nu = 15\;{\rm MeV}$, but within 10 MeV and 25 MeV of neutrino energy, the cross section does not vary significantly for the chosen $m_\chi $ and $m_\phi$. Also, for this choice of masses our forecasted bound is stronger than the cluster and laboratory constraints.
In our case, the expected number of neutrino events in Hyper-Kamiokande in the detector is significantly larger, which allows us to reach unexplored parameter space with the echo method.

\section{Conclusions}
We have shown that in the event of the next Galactic SN, we can constrain neutrino-DM coupling by looking for the delayed neutrino signal from MeV neutrinos echoing off the local DM in the Galaxy. Depending on the model parameters, the signal can be spread out over a duration of $\mathcal{O}(10^8)$ s. 
For fermionic DM-neutrino interaction via a vector mediator, we can constrain the effective mediator coupling to $g\lesssim 1$ for $\sim10-100$~keV DM and $\mathcal{O}(10)$~MeV mediators. In this model, the bounds from our work are more stringent than those from cluster constraints for $m_\chi<20$ keV. We lose the ability to constrain mediators masses, $m_V\lesssim 100$~eV, where most of the delayed signal is contained in a time window shorter than the duration of the neutrino burst.
For fermionic DM and a scalar mediator, constraints for $m_\chi\lesssim 1$~keV are stronger than other bounds for $m_\phi$ between 1 and 20 MeV. Above this DM mass, cluster constraints are stronger for $g_\nu/g_\chi$ ratios consistent with $g_\nu<0.1$.
For scalar DM and a scalar mediator, constraints can be better than cluster constraints for $\sim10-100$~keV DM and $\mathcal{O}(10)$~MeV mediators, provided that we adjust the $g_\nu/g_\chi$ ratio accordingly. In this model, however, the laboratory bound on $g_\nu$ becomes much stronger for the large $g_\nu/g_\chi\gtrsim 10^5-10^7$ ratios used, such that only $m_\phi\lesssim$ 2 MeV can be explored for DM masses below 20 keV.

This study has presented a novel approach to probe DM-neutrino interaction with MeV neutrinos from SNe. The neutrino echo method may access the parameter space that have not been explored by DM direct detection searches due to their energy threshold or cosmology. Next-generation neutrino detectors such as Hyper-Kamiokande and DUNE as well as JUNO would be able to explore the keV-MeV DM region due to the large number of expected SN neutrino interactions in these detectors.
\\

{\bf Acknowledgements.} 
We would like to thank Matheus Hostert and Sergio Palomares-Ruiz for useful comments and discussions. We also thank an anonymous referee for helpful comments that improved the manuscript. 
The authors acknowledge the support from Kavli Institute for Theoretical Physics. This research was supported in part by the National Science Foundation under Grant No.~NSF PHY-1748958.
J.C. is supported by the NSF Grant No.~AST-1908689 and AST-2108466.
A.K. acknowledges the support from the Institute for the Gravitation and the Cosmos through IGC postdoctoral fellowship award. The work of K.M. is supported by the NSF Grant No.~AST-1908689, No.~AST-2108466 and No.~AST-2108467, and KAKENHI No.~20H01901 and No.~20H05852. 

\bibliographystyle{JHEP}
\bibliography{nubib,bibtex}

\end{document}